\documentclass[twocolumn]{aastex701}
\usepackage{subcaption}
\usepackage{amsmath}
\usepackage{bm}

\defcitealias{Zou+2025}{Z25}

\begin{document}

\title{The black hole occupation fraction as a fossil record of seeding, dynamics, and galaxy assembly}

\author[orcid=0009-0000-6612-0599]{Emma Jane Weller}
\affiliation{Department of Astronomy, Yale University, New Haven, CT 06511, USA}
\email[show]{emma.weller@yale.edu}

\author[0000-0002-5554-8896]{Priyamvada Natarajan}
\affiliation{Department of Astronomy, Yale University, New Haven, CT 06511, USA}
\affiliation{Department of Physics, Yale University, New Haven, CT 06511, USA}
\affiliation{Black Hole Initiative, Harvard University, Cambridge, MA 02138, USA}
\email{}

\author[0000-0001-9947-6911]{Colin J. Burke}
\affiliation{Department of Astronomy, Yale University, New Haven, CT 06511, USA}
\affiliation{Department of Physics, University of North Texas, Denton, TX 76203, USA}
\email{}
 
\begin{abstract}
The black hole occupation fraction (BHOF) is a powerful but intrinsically layered diagnostic of massive black hole (BH) seeding and subsequent galaxy assembly. We measure the total, central, and wandering BHOF in the ASTRID cosmological hydrodynamical simulation from $z=5$ to $z=0$ in galaxies with stellar masses ranging from $10^7$--$10^{12} \, \mathrm{M}_\odot$. For the full population, the total BHOF remains close to unity across most stellar masses and redshifts, reflecting efficient seeding in eligible halos. At low stellar masses, the central BHOF declines toward the present day, while the wandering BHOF rises, indicating that mergers deposit BHs at off-center locations in galaxies. Compared to these full-population BHOFs, the heavy-seed BHOFs are substantially smaller at the low-mass end. The active BHOFs are smaller at all stellar masses, especially at late times, demonstrating that active galactic nuclei (AGN)-selected samples trace duty cycles rather than intrinsic BH occupation. At $z=0$, primary (central) galaxies have higher BHOFs than satellites, and star-forming low-mass galaxies preferentially host wandering rather than central BHs. Our results show that the BHOF, when decomposed by BH location, seeding history, accretion state, and larger-scale galactic environment, encodes a rich fossil record of BH origins and dynamics. 
\end{abstract}


\section{Introduction} \label{sec:intro}

The ubiquity of massive black holes (BHs) at the centers of massive galaxies is now one of the empirical cornerstones of galaxy evolution. Dynamical measurements in the local Universe, reverberation mapping of active galactic nuclei (AGN), and the empirically derived tight correlations between central BH mass and host galaxy properties all point to a close connection between the assembly of galaxies and the growth of their central BHs (e.g., \citealt{Haehnelt+1998,Magorrian1998,KormendyHo2013,Greene2020}). Yet this apparently orderly picture is anchored primarily at the high-mass end, where the black hole occupation fraction (BHOF) is expected to be close to unity. At lower stellar masses, the BHOF proves to be profoundly diagnostic.

We define the BHOF as the fraction of galaxies of a given stellar mass that contain at least one massive BH. While the BH mass function and AGN luminosity function are more strongly reshaped by accretion, mergers, and feedback, the BHOF is expected to retain a sharper memory of the initial conditions under which the first BHs formed. Consequently, different seeding mechanisms produce sharply different predictions for the BHOF \citep{RicarteNatarajan2018}. Light seeds, plausibly produced as remnants of Population III stars, are expected to be numerous but initially low-mass, with characteristic masses of $\sim 10^2 \, \mathrm{M}_\odot$ (e.g., \citealt{MadauRees2001,Volonteri2003}). Heavy seeds, produced, for instance, by direct collapse in rare pristine atomic-cooling halos, are expected to be less common but substantially more massive, with initial masses of $\sim 10^4$--$10^6 \, \mathrm{M}_\odot$ (e.g., \citealt{BrommLoeb2003,Begelman2006,LodatoNatarajan2006,NatarajanVolonteri2012,RicarteNatarajan2018,Bhowmick+2026}). These different birth channels converge rapidly in massive galaxies, where hierarchical assembly and efficient growth drive the BHOF toward unity \citep{Menou+2001}. In low-mass galaxy populations, however, the imprint of seeding is anticipated to survive to low redshift.

Observationally, the low-mass regime is also the most difficult to access. Direct dynamical detections require exquisite spatial resolution and are largely restricted to nearby galaxies. AGN-based searches, whether in the optical, radio, X-ray, or infrared, recover only the subset of BHs that are actively accreting above the relevant survey threshold. As such, the measured BHOF is sensitive to accretion physics, duty cycle, obscuration, and selection effects. Nonetheless, sustained observational efforts have begun to place increasingly informative constraints on the local BHOF in dwarf and low-mass galaxies (e.g., \citealt{Greene2012,Miller2015,Burke2023_quasars,Burke2025,Zou+2025}). In particular, X-ray searches provide a comparatively clean route to identifying accreting massive BHs in nearby galaxies, though even these constraints must be interpreted through models for the underlying active fraction and accretion rate distribution \citep{Chadayammuri+2023}.

On the theoretical side, cosmological simulations have emerged as useful tools for interpreting the BHOF. They allow us to follow the joint evolution of galaxies and BHs across cosmic time and investigate the effects of mergers, environment, numerical resolution, and different implementations of seeding, accretion, feedback, and dynamics. Predictions vary greatly across simulations (e.g., \citealt{Habouzit2017,Habouzit2019,Bellovary2019,Haidar+2022,Sharma+2022,DiMatteo2023,Tremmel2024,Bhowmick2025,Contini2026}), showing that the BHOF is sensitive to many of these factors, especially at stellar masses below $\sim 10^9$--$10^{10} \, \mathrm{M}_\odot$. 

An understudied dimension of the BHOF is the location of BHs within their host galaxies, which our work here uniquely addresses. Cosmological assembly produces a population of off-nuclear wandering BHs, especially in dwarf galaxies with shallow potentials and in massive galaxies with rich merger histories (e.g., \citealt{Tremmel2015,Tremmel+2018,Bellovary2019,Ricarte+2021_origins,DiMatteo2023,Weller2023,Weller+2026}). We use the ASTRID cosmological simulation to study the BHOF from $z=5$ to $z=0$ across five decades in stellar mass, $10^7$--$10^{12} \, \mathrm{M}_\odot$. Because ASTRID implements a subgrid dynamical friction prescription rather than artificially pinning BHs to their host galaxy centers \citep{Chen+2022,Ni+2022}, we are able to separately calculate the total, central, and wandering BHOFs. A near-unity total BHOF may coexist with a substantially lower central BHOF, and the difference between the two encodes the dynamical history of galaxy mergers and orbital decay. This distinction is particularly important when comparing simulations to observations, since most current observational searches are mainly sensitive to central, actively accreting BHs rather than the full underlying population, which includes wanderers. \cite{DiMatteo2023} previously computed the total and wandering BHOFs in ASTRID at $z \sim 2$, but our work covers a range of redshifts and investigates the effects of additional BH and galaxy properties. For the first time, we quantify how the total, central, and wandering BHOFs depend on seed mass, accretion state, star formation activity, and environment over $12.6 \, \mathrm{Gyr}$.

This Letter is organized as follows. In Section~\ref{sec:simulation}, we describe the ASTRID simulation and its BH seeding, accretion, feedback, and dynamical prescriptions. In Section~\ref{sec:methods}, we define our galaxy sample, distinguish between central and wandering BHs, and describe the population cuts used throughout the analysis. In Section~\ref{sec:results}, we present the redshift evolution of the total, central, and wandering BHOFs, examine the effects of heavy-seed and active BH selections, and quantify the dependence of the BHOF on galaxy type. We compare our predictions to current observational constraints and to other cosmological simulations. Finally, in Section~\ref{sec:conclusion}, we summarize our key findings and discuss their implications for future observational searches.

\section{The ASTRID simulation} \label{sec:simulation}

ASTRID is a cosmological simulation run from $z=99$ to $z=0$ using the smoothed particle hydrodynamics code MP-Gadget \citep{Feng+2018}. Its cosmological parameters follow \cite{Planck2020}. The simulation contains $5500^3$ cold dark matter particles and an initially equal number of gas particles in a box with a side length of $250 \, h^{-1} \, \mathrm{cMpc}$. The dark matter and initial gas particle masses are $6.74 \times 10^6 \, h^{-1} \, \mathrm{M}_\odot$ and $1.27 \times 10^6 \, h^{-1} \, \mathrm{M}_\odot$, respectively, and the gravitational softening length is $1.5 \, h^{-1} \mathrm{ckpc}$. Halos and subhalos are identified using the friends-of-friends \citep{Davis+1985} and SUBFIND \citep{Springel+2001} algorithms, respectively. 

ASTRID uses halo-based BH seeding over cosmic time. A BH is seeded when a halo that does not yet contain a BH exceeds a total mass of $5 \times 10^9 \, h^{-1} \, \mathrm{M}_\odot$ and a stellar mass of $2 \times 10^6 \, h^{-1} \, \mathrm{M}_\odot$. The densest gas particle in the halo is then converted into a BH particle, which inherits the position and velocity of the parent gas particle. The seed mass is drawn from a power-law probability distribution with a minimum of $3 \times 10^4 \, h^{-1} \, \mathrm{M}_\odot$, a maximum of $3 \times 10^5 \, h^{-1} \, \mathrm{M}_\odot$, and a power-law index of $-1$. The accretion of gas onto the BH is estimated using a Bondi-Hoyle-Lyttleton-like prescription \citep{DiMatteo+2005}. AGN feedback operates in either a high-accretion quasar mode or a low-accretion radio (jet) mode, which injects energy into the surrounding gas as thermal or kinetic feedback. At $z > 2.3$, BHs operate only in the high-accretion mode, but at lower redshifts, massive and minimally accreting BHs can enter the low-accretion mode. Unlike many cosmological simulations, ASTRID does not pin BHs to the centers of their hosts, instead applying a subgrid dynamical friction model \citep{Tremmel2015, Chen+2022}. This gives rise to a population of BHs wandering off-center in their host galaxies \citep{DiMatteo2023,Weller2023,Weller+2026}. We note, however, that ASTRID does not include gravitational-wave recoil or small-scale three-body interactions, and therefore does not contain any off-center BHs produced by these processes. Two BHs merge when their separation is less than twice the gravitational softening length and their kinetic energy has dissipated enough that they are gravitationally bound. For more detailed descriptions of ASTRID, including the additional subgrid models employed, we refer readers to the simulation presentation papers: \cite{Bird+2022, Ni+2022, Ni+2025, Zhou+2026}.

\section{Methods} \label{sec:methods}

We began by identifying all galaxies (corresponding to SUBFIND subhalos) with stellar masses between $10^7 \, \mathrm{M}_\odot$ and $10^{12} \, \mathrm{M}_\odot$ at six ASTRID snapshots: $z = 0,1,2,3,4,5$. We defined central BHs as those within the stellar half-mass radius of their host galaxy, and wandering BHs as those outside of this radius. Given the minimum seed mass in ASTRID, all BH particles are massive BHs with $M_\mathrm{BH} \geq 3 \times 10^4 \, h^{-1} \, \mathrm{M}_\odot$. We computed the central, wandering, and total (central+wandering) BHOFs: the fraction of galaxies, as a function of stellar mass, that contain at least one central BH, at least one wandering BH, or at least one of either, respectively. We used 50 evenly spaced stellar mass bins and excluded bins containing fewer than 100 galaxies. Within these bins, we also computed the median and the 5th to 95th percentile range of the total number of BHs (centrals+wanderers).

We defined a BH as a heavy-seed BH if its seed mass, or the seed mass of any BH that had previously merged with it, exceeded $10^5 \, \mathrm{M}_\odot$. We considered BHs with an Eddington ratio (accretion rate divided by the Eddington rate, $f_\mathrm{Edd} = \dot{M}_\mathrm{BH} / \dot{M}_\mathrm{Edd}$) above $0.01$ to be active. In each halo, we labeled the galaxy with the largest number of bound particles (generally equivalent to the central galaxy in a galaxy group) as the primary, and the others as satellites. We computed the specific star formation rate (star formation rate per unit stellar mass, $\mathrm{sSFR} = \mathrm{SFR} / M_\star$) of the galaxies and classified them as star-forming ($\mathrm{sSFR} > 10^{-10} \, \mathrm{yr}^{-1}$) or quiescent ($\mathrm{sSFR} \leq 10^{-10} \, \mathrm{yr}^{-1}$).

Throughout this Letter, we use the term ``full population'' to indicate that we have not applied any selections based on the BH and galaxy properties described in the previous paragraph (heavy-seed, active, primary/satellite, star-forming/quenched). In the following Section~\ref{sec:results}, we present the central, wandering, and total (central+wandering) BHOFs for the full population and several sub-populations.

\section{Results} \label{sec:results}

\subsection{Overall trends} \label{sec:overall}

\begin{figure}
    \centering
    \includegraphics[width=1\columnwidth]{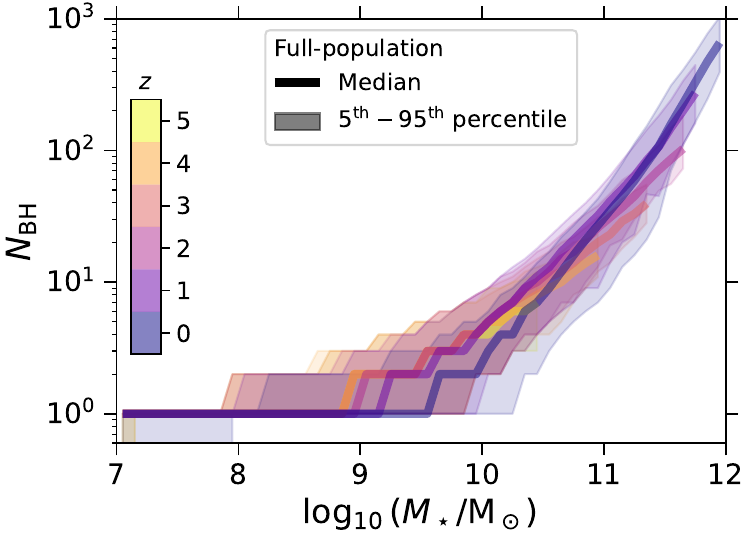}
    \caption{Total number of black holes per galaxy ($N_\mathrm{BH}$), as a function of stellar mass, for the full population from $z=5$ to $z=0$. Solid lines show the median in each stellar mass bin, while shaded regions indicate the 5th--95th percentile range. Low-mass galaxies typically host either zero or one BH. The characteristic stellar mass threshold, above which galaxies typically host multiple BHs, shifts to higher masses at later times. This reflects continued stellar mass growth while the formation of new BH seeds declines. In massive systems, $N_\mathrm{BH}$ rises steeply with stellar mass due to mergers that build up the BH population.}
\label{fig:Nbh}
\end{figure}

\begin{figure}
    \centering
    \includegraphics[width=1\columnwidth]{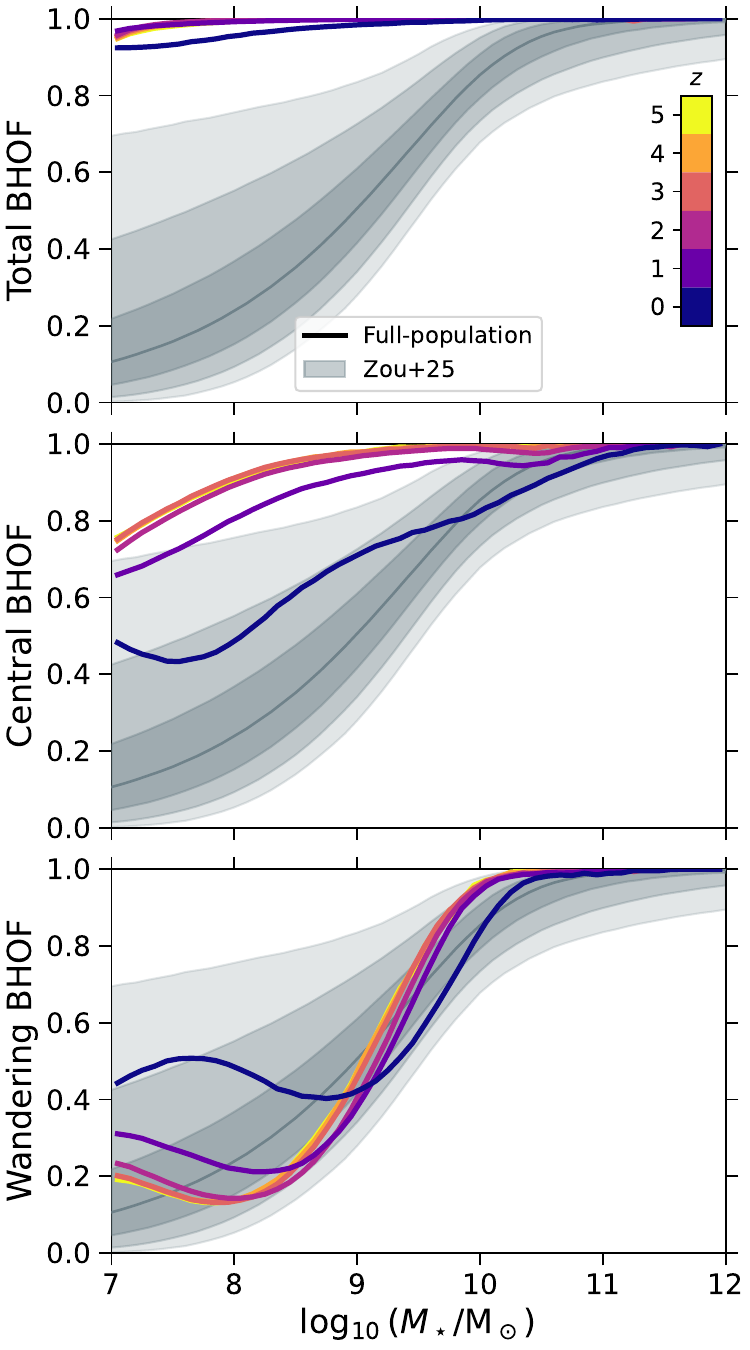}
    \caption{Total, central, and wandering BHOFs for the full population from $z=5$ to $z=0$. The total BHOF remains close to unity across the studied stellar mass and redshift ranges, indicating that most galaxies in the sample host at least one BH. However, the balance between central and wandering BHs evolves strongly: at low stellar masses, the central BHOF decreases toward $z=0$, while the wandering BHOF increases. This demonstrates that late-time low-mass galaxies can remain occupied even when their BHs are displaced from the galaxy center. Gray bands show local observational constraints from \citetalias{Zou+2025}, with a solid line marking the median and shaded regions indicating the 1$\sigma$, 2$\sigma$, and 3$\sigma$ intervals.}
    \label{fig:all}
\end{figure}

We begin by considering Figure~\ref{fig:Nbh}, which shows the typical number of BHs per galaxy ($N_\mathrm{BH}$) as a function of galaxy stellar mass ($M_\star$) from $z=5$ to $z=0$. At the low end of our studied stellar mass range, most galaxies contain exactly one BH, but some have none (for $M_\star < 10^8 \, \mathrm{M}_\odot$, $7\%$ of $z=0$ galaxies have no BHs and $92\%$ have one; at higher redshifts, $2$--$3\%$ of galaxies have no BHs and $96$--$97\%$ have one). At the high-mass end, nearly all galaxies contain one or more BHs (for $M_\star > 10^8 \, \mathrm{M}_\odot$, $98\%$ of $z=0$ galaxies and $>99\%$ of higher-redshift galaxies have at least one BH).

The stellar mass at which the median number of BHs increases to two, three, four, and so on is higher at lower redshifts. This is because low-$M_\star$ galaxies grow primarily via star formation, which increases $M_\star$ but not $N_\mathrm{BH}$, pushing features toward larger $M_\star$ as time progresses. At late times, very few new BHs are formed. However, at the high-$M_\star$ end, the slope of the curve increases with time. This can be attributed to galaxy mergers, which lead to increases in $M_\star$ and even more rapid jumps in $N_\mathrm{BH}$.

In Figure~\ref{fig:all}, we plot the total, central, and wandering BHOFs for the full population from $z=5$ to $z=0$. In general, the BHOF tends to increase with stellar mass, as expected since larger galaxies are more likely to host massive BHs. The total full-population BHOF is always close to unity, but it is slightly lower at $z=0$ than at higher redshifts. This is consistent with Figure~\ref{fig:Nbh}, where we saw that most galaxies have at least one BH, but some low-$M_\star$ galaxies have none, especially at $z=0$. At the low-$M_\star$ end, the central full-population BHOF decreases with decreasing redshift, while the wandering full-population BHOF increases correspondingly. This inverse relationship is consistent with Figure~\ref{fig:Nbh} as well: most of the low-$M_\star$ galaxies contain exactly one BH, which could be either a central or a wanderer. Wanderers are generally a result of galaxy mergers \citep{Ricarte+2021_origins,Weller+2026}, which likely explains their increased presence at later times. The decrease in the central BHOF with decreasing redshift has also been seen in previous studies \citep{Haidar+2022,Sharma+2022,Tremmel2024}. The origin of the bump at $10^7 \lesssim M_\star / M_\odot \lesssim 10^8$ in the $z=0$ wandering full-population BHOF (and the corresponding dip in the central BHOF) is unclear. As mentioned in Section \ref{sec:intro}, wandering BHs are common in galaxies with rich merger histories and dwarfs with shallow potentials. The bump may correspond to a regime where these effects combine to produce a particularly high fraction of massive BHs that are off-center.

We compare our findings to local ($z \sim 0$) constraints on the central BHOF from \citeauthor{Zou+2025} (\citeyear{Zou+2025}; hereafter Z25), based on Chandra X-ray imaging of 1606 galaxies within $50 \, {\rm Mpc}$. They provide the median and the 1$\sigma$, 2$\sigma$, and 3$\sigma$ ranges of their sampling results. Their observations probe Eddington ratios well below the AGN regime, allowing them to estimate the central BHOF for the full BH population, not just active BHs. They constrain the low-mass end of the BHOF to significantly lower values than previous studies. Promisingly, our central full-population BHOF is consistent with their $3\sigma$ constraints at $z=0$.

In the following Sections \ref{sec:BH_props} and \ref{sec:gal_props}, we study subsets of our full population to understand how the BHOF depends on the selection criteria applied to the BHs and galaxies.

\begin{figure*}
    \centering
    \includegraphics[width=2\columnwidth]{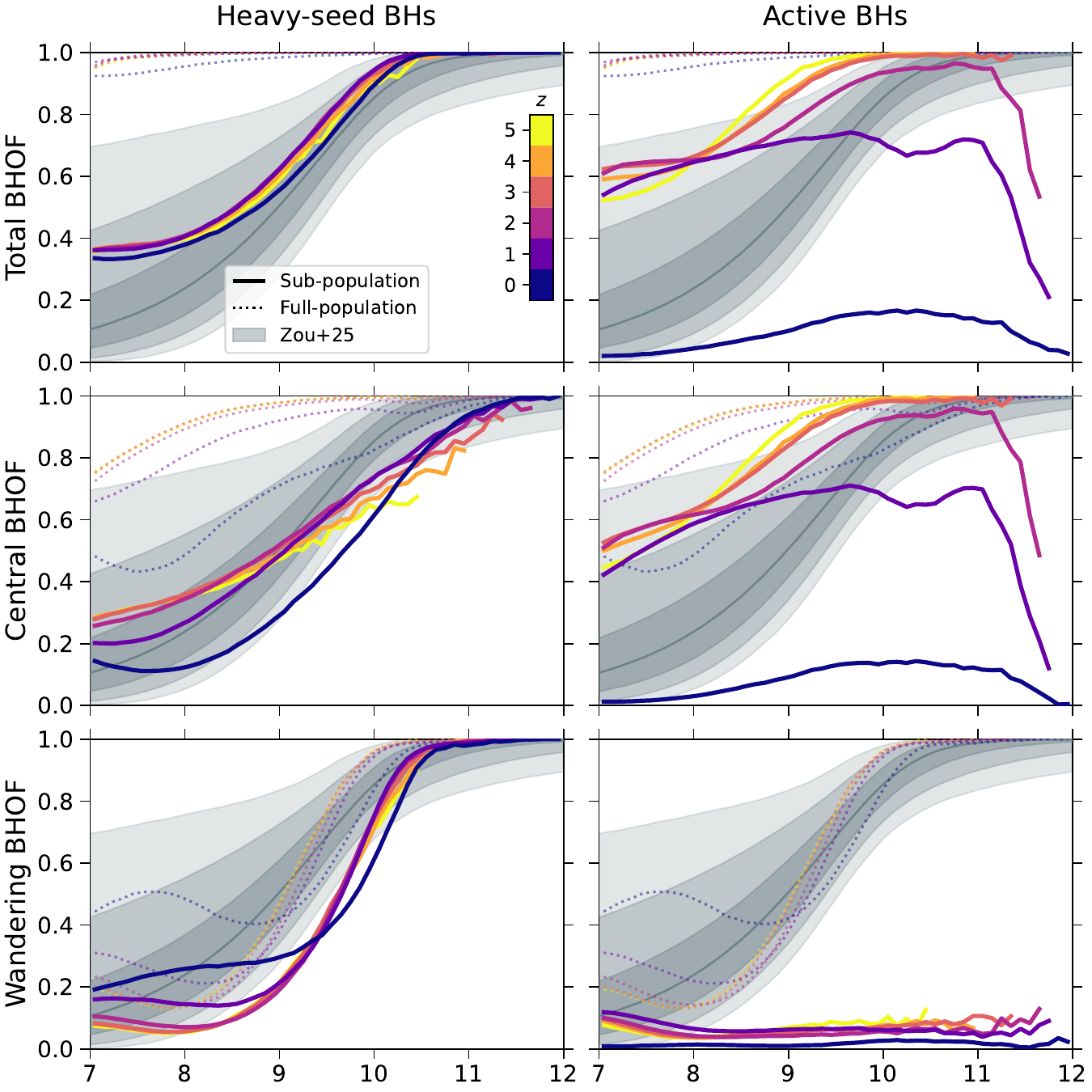}
    \caption{Total, central, and wandering BHOFs from $z=5$ to $z=0$ for heavy-seed BHs (thick lines, left column) and active BHs (thick lines, right column). We define heavy-seed BHs as BHs for which the initial seed mass, or the seed mass of any progenitor that merged into them, exceeded $10^5 \, \mathrm{M}_\odot$. We define active BHs as those with $f_{\rm Edd} > 0.01$. Thin dotted lines mark the corresponding full-population BHOFs, while gray bands show local observational constraints from \citetalias{Zou+2025}. The heavy-seed BHOFs are substantially lower than the full-population BHOFs at low stellar masses, demonstrating that the low-mass end of the BHOF is sensitive to the underlying seed mass distribution. For active BHs, the total and central BHOFs decline strongly toward low redshift, particularly after cosmic noon, while the wandering BHOF is negligible at nearly all masses and redshifts. This shows that AGN-selected samples trace the accretion duty cycle rather than the intrinsic BHOF.}
    \label{fig:bhs}
\end{figure*}

\subsection{BH properties} \label{sec:BH_props}

In the left column of Figure~\ref{fig:bhs}, we show the BHOFs for heavy-seed BHs (seed mass $>10^5 \, \mathrm{M}_\odot$, as described in Section~\ref{sec:methods}). The heavy-seed BHOFs are lower than their full-population counterparts, especially at the low-mass end. As a result, the total heavy-seed BHOF falls within the $2\sigma$ constraints from \citetalias{Zou+2025} at all redshifts. However, the central heavy-seed BHOF is slightly lower than the \citetalias{Zou+2025} constraints at intermediate stellar masses.

In the right column of Figure~\ref{fig:bhs}, we show the BHOFs for active BHs ($f_\mathrm{Edd} > 0.01$, as described in Section~\ref{sec:methods}). The wandering active BHOF is very small for all stellar masses and redshifts, which is expected, as wandering BHs generally experience minimal accretion \citep{Weller2023}. The total and central active BHOFs (which are essentially the same) decrease significantly with decreasing redshift, especially after cosmic noon ($z \sim 2$). Notably, while the total and central active BHOFs fall within or above the \citetalias{Zou+2025} local constraints at $z \gtrsim 1$, they drop well below these observational bounds at $z=0$. The active BHOFs fall off at the high-$M_\star$ end, likely because the massive central BHs have expelled gas from their surroundings via AGN feedback and therefore have little gas left to accrete.

\subsection{Galaxy properties} \label{sec:gal_props}

\begin{figure}
    \centering
    \includegraphics[width=1\columnwidth]{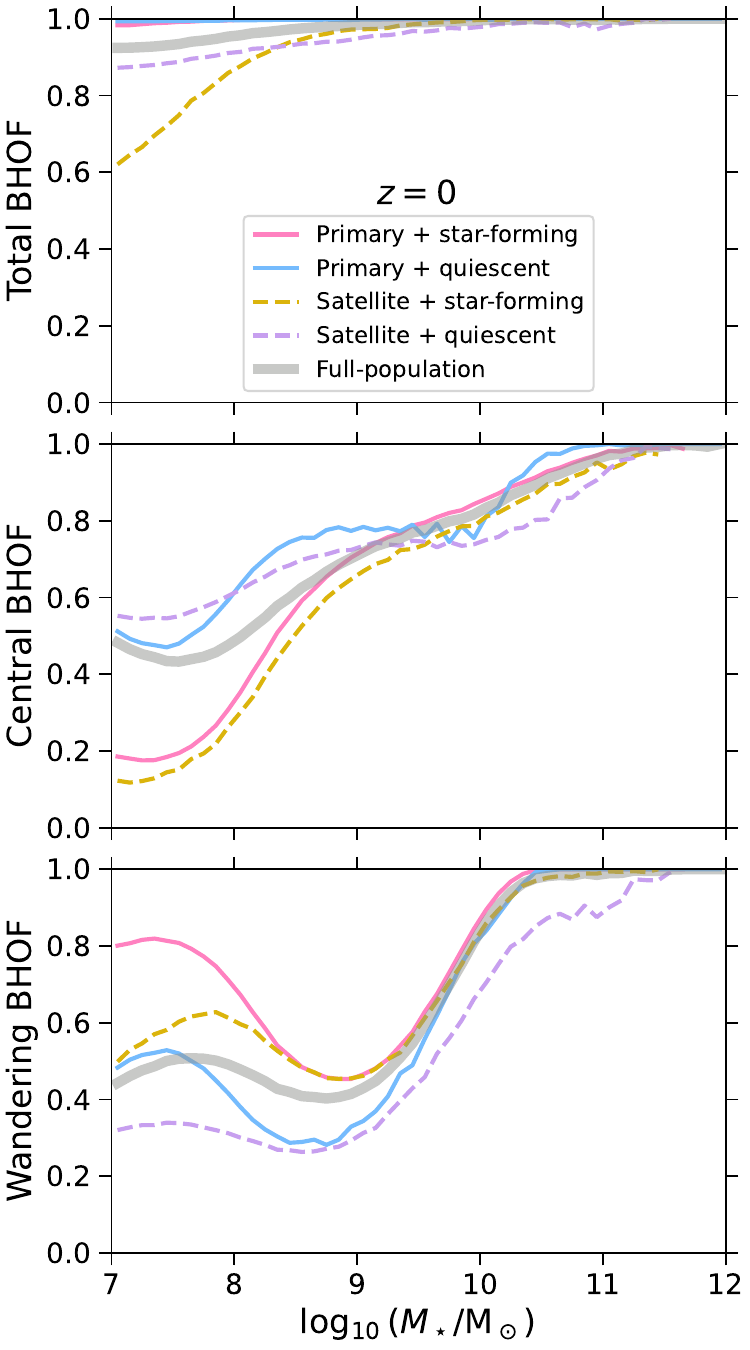}
    \caption{Dependence of the $z=0$ total, central, and wandering BHOFs on galaxy environment and star formation state. We compare primary and satellite galaxies, each divided into star-forming and quiescent populations using ${\rm sSFR} = 10^{-10} \, \mathrm{yr}^{-1}$ as the threshold. The $z=0$ full-population BHOFs are shown for reference. Primary galaxies generally have higher BHOFs than satellites, consistent with preferential BH seeding and accumulation in the densest regions of halos. At low stellar masses, quiescent galaxies have higher central BHOFs and lower wandering BHOFs than star-forming galaxies, suggesting that the location of the BH within the host is correlated with the galaxy's star formation state.}
\label{fig:gals}
\end{figure}

In Figure~\ref{fig:gals}, we plot the BHOFs for different types of galaxies. We show only the $z=0$ BHOFs, where the changes relative to the full-population BHOFs are most significant. Star-forming/quiescent \textit{primary} galaxies generally have higher total, central, and wandering BHOFs than star-forming/quiescent \textit{satellite} galaxies. This is expected, given that BHs are seeded in the densest regions of halos, and galaxy mergers bring BHs into primary galaxies. At the low-$M_\star$ end, primary/satellite \textit{star-forming} galaxies have lower central BHOFs and higher wandering BHOFs than primary/satellite \textit{quiescent} galaxies, which also makes sense: the low-$M_\star$ galaxies usually have at most one BH, and a central is more likely than a wanderer to quench its host via AGN feedback. For primary galaxies, the total BHOF is near $100\%$ at all stellar masses, but it falls to $\sim 90\%$ for quiescent satellites and $\sim 60\%$ for star-forming satellites at the lowest stellar masses.

\subsection{Comparison to previous results}

\begin{figure*}
    \centering
    \includegraphics[width=1.5\columnwidth]{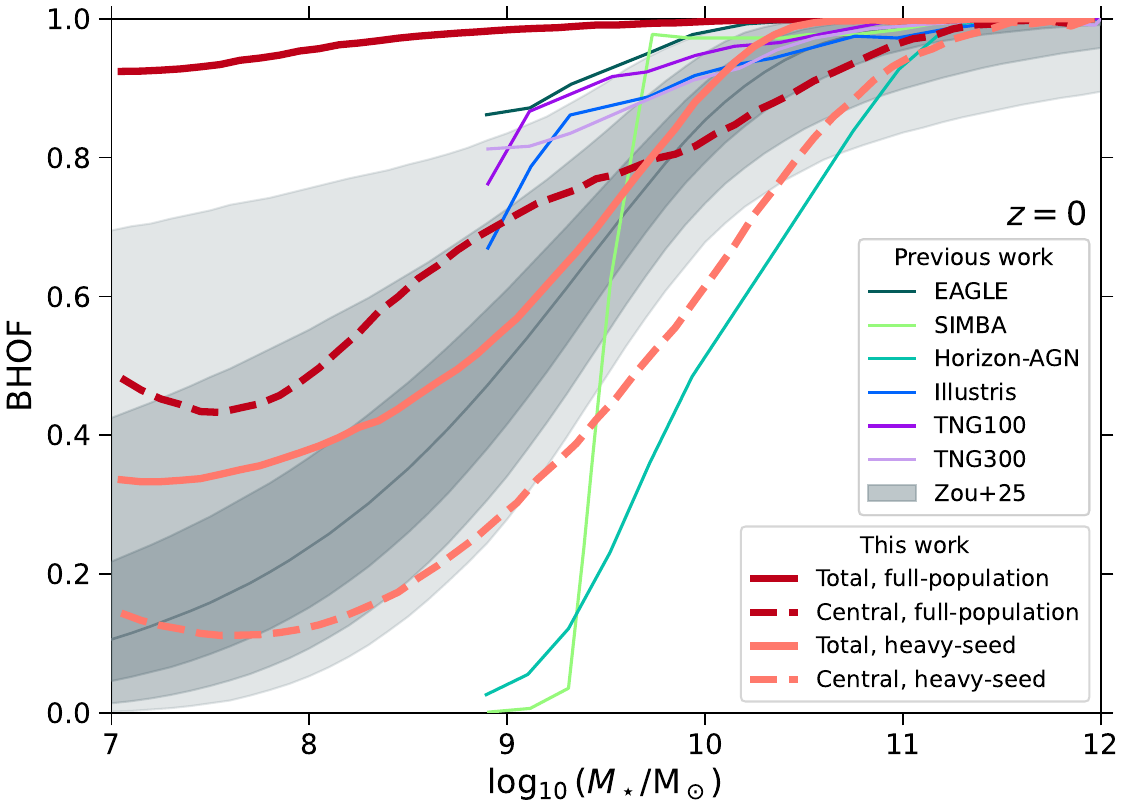}
    \caption{Comparison of $z=0$ ASTRID BHOFs with observational constraints and predictions from other simulations. We plot the ASTRID total and central full-population BHOFs, as well as the corresponding heavy-seed BHOFs. Local observational constraints from \citetalias{Zou+2025} are shown alongside predictions from the EAGLE, SIMBA, Horizon-AGN, Illustris, TNG100, and TNG300 cosmological simulations, computed by \cite{Haidar+2022}. The wide spread among simulation results, especially at $M_\star \lesssim 10^9$--$10^{10} \, \mathrm{M}_\odot$, shows that the low-mass end of the BHOF is sensitive to variations between the simulations (e.g., BH seeding, feedback, and dynamics models, numerical resolution, and box size).}
\label{fig:comparison}
\end{figure*}

In Figure~\ref{fig:comparison}, we plot the ASTRID $z=0$ total and central BHOFs for the full population and for heavy-seed BHs, and compare our results to the $z=0$ central BHOFs from six other cosmological simulations (EAGLE, SIMBA, Horizon-AGN, Illustris, TNG100, and TNG300), computed for galaxies with $M_\star \geq 10^9 \, \mathrm{M}_\odot$ by \cite{Haidar+2022}. We obtained the BHOFs from Figure~1 in their paper using WebPlotDigitizer \citep{WebPlotDigitizer}. While all of the plotted BHOFs are near unity at high stellar masses, they diverge at lower masses, illustrating the sensitivity of the low-mass end of the BHOF to variations between the simulations, such as the BH seeding, feedback, and dynamics models and simulation volume and resolution. We note that the abrupt jump in the SIMBA BHOF can be attributed in part to the fact that this simulation does not seed BHs in galaxies with $M_\star < 10^{9.5} \, \mathrm{M}_\odot$. We also show the local observational constraints from \citetalias{Zou+2025}. Since all the simulations have near-unity BHOFs at the high-mass end, they are all consistent with observations at high stellar masses. As noted in previous sections, ASTRID shows consistency with observations at the low-mass end as well: the central full-population BHOF, central heavy-seed BHOF, and total heavy-seed BHOF fall within the \citetalias{Zou+2025} $3\sigma$ constraints.

\section{Discussion and conclusions} \label{sec:conclusion}

The main conclusions of our work are as follows:
\begin{enumerate}
    \item \textbf{The total full-population BHOF in ASTRID is near unity across the studied stellar mass range ($\bm{10^7 < M_\star/\mathrm{M}_\odot < 10^{12}}$) and redshift range ($\bm{0 \leq z \leq 5}$).} This reflects the efficiency of the ASTRID seeding prescription: BHs are seeded in halos once they exceed the adopted total and stellar mass thresholds, and the resulting BH population is then propagated forward through hierarchical assembly. The BHOF is significantly more informative when it is computed for subsets of the BH and galaxy population.
    \item \textbf{At low stellar masses ($\bm{M_\star / \mathrm{M}_\odot \lesssim 10^9}$--$\bm{10^{10}}$), the central BHOF tends to decrease over cosmic time, while the wandering BHOF increases.} This is a consequence of hierarchical assembly: mergers can perturb central BHs and deposit off-nuclear BHs, which can remain off-center for long timescales due to inefficient dynamical friction, especially in dwarf galaxies. Most low-mass systems contain no more than one massive BH, so as the wandering BHOF increases, the central BHOF decreases correspondingly. Thus, a galaxy can remain occupied in the total sense while no longer hosting a central BH. This distinction is essential for interpreting observations because most current searches are primarily sensitive to central and/or actively accreting BHs. A low central BHOF may not imply inefficient BH seeding; it could instead indicate that BHs were seeded efficiently but subsequently displaced or delivered to the outskirts of galaxies where dynamical friction is inefficient at sinking them. The difference between the total and central BHOF can therefore reveal information about the dynamical evolution of the BH population after seeding.
    \item \textbf{Selecting only the descendants of heavy BH seeds ($\bm{M_\mathrm{BH} > 10^5 \, \mathrm{M}_\odot}$) results in BHOFs that are substantially smaller than their full-population counterparts, especially at low stellar masses.} This shows that the low-mass end of the BHOF remains a probe of the initial seed mass distribution even after subsequent growth and mergers. The heavy-seed BHOF in ASTRID is closer to current observational constraints than the full-population BHOF over much of the low-mass regime.
    \item \textbf{The BHOF for actively accreting BHs ($\bm{f_\mathrm{Edd} > 0.01}$) is a poor proxy for the full-population BHOF, especially at low redshift.} The total and central active BHOFs are dramatically lower than their full-population counterparts, and display a particularly strong decline after cosmic noon and at high stellar masses. This reflects the transition from rapid growth at earlier epochs to feedback-regulated, gas-poor evolution, starting with the most massive BHs. The wandering active BHOF is near-zero, consistent with the expectation that wanderers typically accrete at low levels due to their low-density environments. Our results emphasize that translating the BHOF measured from AGN-selected samples into an intrinsic, full-population BHOF requires a physically calibrated model for the active fraction.
    \item \textbf{The BHOF depends on the environment and star formation state of the galaxies in the sample: at $z=0$, primary galaxies tend to have higher BHOFs than satellites, and low-mass quiescent galaxies have higher central BHOFs (and correspondingly lower wandering BHOFs) than their star-forming counterparts.} The first point is consistent with the expectation that BHs are preferentially seeded in the densest regions of halos and subsequently accumulated by primary galaxies through mergers. The second point suggests that a centrally located BH may be more capable of coupling energetically to the surrounding medium and regulating the central gas reservoir. We do not claim from this analysis alone that BH feedback is the sole driver of quenching in low-mass galaxies; however, the correlation between quiescence and central occupation suggests that the location of a BH, and not merely its presence, may impact the state of the host galaxy.
\end{enumerate}

The high-mass end of the BHOF is largely saturated: massive galaxies almost invariably host BHs, and therefore their occupation fractions carry limited discriminatory power. At lower masses, however, the BHOF is highly informative. The imprint of different seeding prescriptions is more pronounced, mergers can displace BHs from galaxy centers, accretion is intermittent, and the environment can strongly impact occupation statistics. This is reflected in comparisons between simulations: the predicted BHOFs vary widely at low stellar masses. As observational constraints improve, especially at $M_\star \lesssim 10^9$--$10^{10} \, \mathrm{M}_\odot$, the BHOF will become an increasingly powerful tool. Promisingly, the ASTRID $z=0$ central BHOF is consistent with local observational constraints from \citetalias{Zou+2025} for the full population and (at most stellar masses) the heavy-seed subset. An important limitation, however, is that ASTRID's minimum BH seed mass prevents us from exploring light seeding below $3 \times 10^4 \, h^{-1} \, \mathrm{M}_\odot$.

In a recent study, \cite{Contini2026} used the semi-analytic model FEGA25 to investigate how the BHOF depends on the adopted BH mass threshold, galaxy type, and redshift. They warn that these trends are also impacted by the simulation volume and resolution, so robust predictions require sufficient resolution to study low-mass galaxies and sufficient volume to sample a wide variety of cosmic environments. We argue that ASTRID meets these requirements: it is a self-consistent cosmological simulation that has a box size comparable to the largest simulations from \cite{Contini2026} (YS200 and YS300), and a dark matter particle mass smaller than that of their high-resolution run (YS50HR). Furthermore, as noted previously, the ASTRID results are broadly consistent with current observational constraints, and our redshift evolution trends agree with previous studies.

In summary, this work shows that the BHOF is most diagnostic when decomposed by BH location, seed history, accretion state, and host galaxy properties. Our approach allows us to disentangle and gain insight into the effects of BH formation mechanisms, accretion histories, dynamics, and environments. Location is a particularly important but historically understudied factor. Separating central and wandering BHs reveals that the dynamical redistribution of BHs within galaxies and halos leaves an imprint on the late-time BHOF, especially at low stellar masses. Future observational progress will come from combining complementary search channels, such as deep X-ray observations of nearby low-mass galaxies, optical and infrared spectroscopic searches for weak AGN, radio constraints on compact accretion, time-domain searches for tidal disruption events, and eventually, dynamical measurements with next-generation facilities, which will each probe different regions of BH parameter space. Theoretical predictions must therefore move beyond a single intrinsic BHOF and toward forward-modeled occupation fractions matched to each observational selection. The framework presented here provides a step in that direction. 

\begin{acknowledgements}

We thank the anonymous referee for constructive comments that improved this Letter. E.J.W. acknowledges support from the National Science Foundation (NSF) Graduate Research Fellowship Program under grant No. DGE-2139841. Any opinions, findings, conclusions, or recommendations expressed in this material are those of the authors and do not necessarily reflect the views of the NSF. P.N. acknowledges support from the Gordon and Betty Moore Foundation and the John Templeton Foundation, which fund the Black Hole Initiative (BHI) at Harvard University, where she is a PI. P.N. also acknowledges support from STScI/NASA via grant JWST-GO-03293024. C.J.B. is supported by an NSF Astronomy and Astrophysics Postdoctoral Fellowship under award AST-2303803. This research award is partially funded by a generous gift of Charles Simonyi to the NSF Division of Astronomical Sciences. The award is made in recognition of significant contributions to Rubin Observatory’s Legacy Survey of Space and Time.
\end{acknowledgements}


\software{Astropy \citep{astropy:2013, astropy:2018, astropy:2022}, Matplotlib \citep{Hunter_2007}, NumPy \citep{Harris+2020}, pandas \citep{McKinney2010, pandas_development_team_2024}, SciPy \citep{Virtanen+2020}, WebPlotDigitizer \citep{WebPlotDigitizer}}

\bibliography{sample701}{}

@article{Magorrian1998,
  author = {{Magorrian}, John and {Tremaine}, Scott and {Richstone}, Douglas and {Bender}, Ralf and {Bower}, Gary and {Dressler}, Alan and {Faber}, S.~M. and {Gebhardt}, Karl and {Green}, Richard and {Grillmair}, Carl and {Kormendy}, John and {Lauer}, Tod},
  title = "{The Demography of Massive Dark Objects in Galaxy Centers}",
  journal = {\aj},
  year = {1998},
  volume = {115},
  number = {6},
  pages = {2285--2305},
  doi = {10.1086/300353}
}

@article{KormendyHo2013,
  author = {{Kormendy}, John and {Ho}, Luis C.},
  title = "{Coevolution (Or Not) of Supermassive Black Holes and Host Galaxies}",
  journal = {\araa},
  year = {2013},
  volume = {51},
  pages = {511--653},
  doi = {10.1146/annurev-astro-082708-101811},
  archivePrefix = {arXiv},
  eprint = {1304.7762},
  primaryClass = {astro-ph.CO}
}

@article{Greene2020,
  author = {{Greene}, Jenny E. and {Strader}, Jay and {Ho}, Luis C.},
  title = "{Intermediate-Mass Black Holes}",
  journal = {\araa},
  year = {2020},
  volume = {58},
  pages = {257--312},
  doi = {10.1146/annurev-astro-032620-021835},
  archivePrefix = {arXiv},
  eprint = {1911.09678},
  primaryClass = {astro-ph.GA}
}

@article{MadauRees2001,
  author = {{Madau}, Piero and {Rees}, Martin J.},
  title = "{Massive Black Holes as Population III Remnants}",
  journal = {\apjl},
  year = {2001},
  volume = {551},
  number = {1},
  pages = {L27--L30},
  doi = {10.1086/319848},
  archivePrefix = {arXiv},
  eprint = {astro-ph/0101223}
}

@article{Volonteri2003,
  author = {{Volonteri}, Marta and {Haardt}, Francesco and {Madau}, Piero},
  title = "{The Assembly and Merging History of Supermassive Black Holes in Hierarchical Models of Galaxy Formation}",
  journal = {\apj},
  year = {2003},
  volume = {582},
  number = {2},
  pages = {559--573},
  doi = {10.1086/344675},
  archivePrefix = {arXiv},
  eprint = {astro-ph/0207276}
}

@article{BrommLoeb2003,
  author = {{Bromm}, Volker and {Loeb}, Abraham},
  title = "{Formation of the First Supermassive Black Holes}",
  journal = {\apj},
  year = {2003},
  volume = {596},
  number = {1},
  pages = {34--46},
  doi = {10.1086/377529},
  archivePrefix = {arXiv},
  eprint = {astro-ph/0212400}
}

@article{Begelman2006,
  author = {{Begelman}, Mitchell C. and {Volonteri}, Marta and {Rees}, Martin J.},
  title = "{Formation of supermassive black holes by direct collapse in pre-galactic haloes}",
  journal = {\mnras},
  year = {2006},
  volume = {370},
  number = {1},
  pages = {289--298},
  doi = {10.1111/j.1365-2966.2006.10467.x},
  archivePrefix = {arXiv},
  eprint = {astro-ph/0602363}
}

@article{LodatoNatarajan2006,
  author = {{Lodato}, Giuseppe and {Natarajan}, Priyamvada},
  title = "{Supermassive black hole formation during the assembly of pre-galactic discs}",
  journal = {\mnras},
  year = {2006},
  volume = {371},
  number = {4},
  pages = {1813--1823},
  doi = {10.1111/j.1365-2966.2006.10801.x},
  archivePrefix = {arXiv},
  eprint = {astro-ph/0606159}
}

@article{NatarajanVolonteri2012,
  author = {{Natarajan}, Priyamvada and {Volonteri}, Marta},
  title = "{The mass function of black holes at $1<z<4.5$: comparison of models with observations}",
  journal = {\mnras},
  year = {2012},
  volume = {422},
  number = {3},
  pages = {2051--2057},
  doi = {10.1111/j.1365-2966.2012.20708.x},
  archivePrefix = {arXiv},
  eprint = {1107.4916},
  primaryClass = {astro-ph.CO}
}

@ARTICLE{RicarteNatarajan2018,
       author = {{Ricarte}, Angelo and {Natarajan}, Priyamvada},
        title = "{The observational signatures of supermassive black hole seeds}",
      journal = {\mnras},
         year = 2018,
        month = dec,
       volume = {481},
       number = {3},
        pages = {3278-3292},
          doi = {10.1093/mnras/sty2448},
archivePrefix = {arXiv},
       eprint = {1809.01177},
 primaryClass = {astro-ph.GA},
       adsurl = {https://ui.adsabs.harvard.edu/abs/2018MNRAS.481.3278R}
}

@ARTICLE{Greene2012,
       author = {{Greene}, Jenny E.},
        title = "{Low-mass black holes as the remnants of primordial black hole formation}",
      journal = {Nature Communications},
         year = 2012,
        month = dec,
       volume = {3},
          eid = {1304},
        pages = {1304},
          doi = {10.1038/ncomms2314},
archivePrefix = {arXiv},
       eprint = {1211.7082},
 primaryClass = {astro-ph.CO},
       adsurl = {https://ui.adsabs.harvard.edu/abs/2012NatCo...3.1304G}
}

@article{Burke2023_quasars,
  author = {{Burke}, C.~J. and {Shen}, Yue and {Liu}, Xin and {Liu}, Tingting and {Loeb}, Abraham and {Ricarte}, Angelo and {Yang}, Qian and {Graham}, Matthew J. and {Djorgovski}, S.~G. and {Mahabal}, Ashish A.},
  title = "{A systematic search for changing-look quasars in ZTF}",
  journal = {\mnras},
  year = {2023},
  volume = {518},
  pages = {1880--1894},
  doi = {10.1093/mnras/stac2478},
  archivePrefix = {arXiv},
  eprint = {2207.04066},
  primaryClass = {astro-ph.GA}
}

@ARTICLE{Burke2025,
       author = {{Burke}, Colin J. and {Natarajan}, Priyamvada and {Baldassare}, Vivienne F. and {Geha}, Marla},
        title = "{Multiwavelength Constraints on the Local Black Hole Occupation Fraction}",
      journal = {\apj},
         year = 2025,
        month = jan,
       volume = {978},
       number = {1},
          eid = {77},
        pages = {77},
          doi = {10.3847/1538-4357/ad94d9},
archivePrefix = {arXiv},
       eprint = {2410.11177},
 primaryClass = {astro-ph.GA},
       adsurl = {https://ui.adsabs.harvard.edu/abs/2025ApJ...978...77B}
}

@ARTICLE{Habouzit2017,
       author = {{Habouzit}, M{\'e}lanie and {Volonteri}, Marta and {Dubois}, Yohan},
        title = "{Blossoms from black hole seeds: properties and early growth regulated by supernova feedback}",
      journal = {\mnras},
         year = 2017,
        month = jul,
       volume = {468},
       number = {4},
        pages = {3935-3948},
          doi = {10.1093/mnras/stx666},
archivePrefix = {arXiv},
       eprint = {1605.09394},
 primaryClass = {astro-ph.GA},
       adsurl = {https://ui.adsabs.harvard.edu/abs/2017MNRAS.468.3935H}
}

@ARTICLE{Habouzit2019,
       author = {{Habouzit}, M{\'e}lanie and {Genel}, Shy and {Somerville}, Rachel S. and {Kocevski}, Dale and {Hirschmann}, Michaela and {Dekel}, Avishai and {Choi}, Ena and {Nelson}, Dylan and {Pillepich}, Annalisa and {Torrey}, Paul and {Hernquist}, Lars and {Vogelsberger}, Mark and {Weinberger}, Rainer and {Springel}, Volker},
        title = "{Linking galaxy structural properties and star formation activity to black hole activity with IllustrisTNG}",
      journal = {\mnras},
         year = 2019,
        month = apr,
       volume = {484},
       number = {4},
        pages = {4413-4443},
          doi = {10.1093/mnras/stz102},
archivePrefix = {arXiv},
       eprint = {1809.05588},
 primaryClass = {astro-ph.GA},
       adsurl = {https://ui.adsabs.harvard.edu/abs/2019MNRAS.484.4413H}
}

@article{Bellovary2019,
  author = {{Bellovary}, J.~M. and {Cleary}, C.~E. and {Munshi}, F. and {Tremmel}, M. and {Christensen}, C.~R. and {Brooks}, A. and {Quinn}, T.~R.},
  title = "{Multimessenger signatures of massive black holes in dwarf galaxies}",
  journal = {\mnras},
  year = {2019},
  volume = {482},
  pages = {2913--2923},
  doi = {10.1093/mnras/sty2842},
  archivePrefix = {arXiv},
  eprint = {1806.00471},
  primaryClass = {astro-ph.GA}
}

@ARTICLE{Haehnelt+1998,
       author = {{Haehnelt}, Martin G. and {Natarajan}, Priyamvada and {Rees}, Martin J.},
        title = "{High-redshift galaxies, their active nuclei and central black holes}",
      journal = {\mnras},
         year = 1998,
        month = nov,
       volume = {300},
       number = {3},
        pages = {817-827},
          doi = {10.1046/j.1365-8711.1998.01951.x},
archivePrefix = {arXiv},
       eprint = {astro-ph/9712259},
 primaryClass = {astro-ph},
       adsurl = {https://ui.adsabs.harvard.edu/abs/1998MNRAS.300..817H}
}

@ARTICLE{Chadayammuri+2023,
       author = {{Chadayammuri}, Urmila and {Bogd{\'a}n}, {\'A}kos and {Ricarte}, Angelo and {Natarajan}, Priyamvada},
        title = "{Constraints From Dwarf Galaxies on Black Hole Seeding and Growth Models With Current and Future Surveys}",
      journal = {\apj},
         year = 2023,
        month = mar,
       volume = {946},
       number = {1},
          eid = {51},
        pages = {51},
          doi = {10.3847/1538-4357/acbea6},
archivePrefix = {arXiv},
       eprint = {2212.04693},
 primaryClass = {astro-ph.GA},
       adsurl = {https://ui.adsabs.harvard.edu/abs/2023ApJ...946...51C}
}

@ARTICLE{Zhou+2026,
       author = {{Zhou}, Yihao and {Di Matteo}, Tiziana and {Bird}, Simeon and {Croft}, Rupert and {Ni}, Yueying and {Yang}, Yanhui and {Chen}, Nianyi and {Lachance}, Patrick and {Zhang}, Xiaowen and {Hafezianzadeh}, Fatemeh},
        title = "{The ASTRID Simulation at z = 0: From Massive Black Holes to Large-scale Structure}",
      journal = {\apj},
         year = 2026,
        month = mar,
       volume = {999},
       number = {1},
          eid = {41},
        pages = {41},
          doi = {10.3847/1538-4357/ae3c08},
archivePrefix = {arXiv},
       eprint = {2510.13976},
 primaryClass = {astro-ph.GA},
       adsurl = {https://ui.adsabs.harvard.edu/abs/2026ApJ...999...41Z}
}

@ARTICLE{Bird+2022,
       author = {{Bird}, Simeon and {Ni}, Yueying and {Di Matteo}, Tiziana and {Croft}, Rupert and {Feng}, Yu and {Chen}, Nianyi},
        title = "{The ASTRID simulation: galaxy formation and reionization}",
      journal = {\mnras},
         year = 2022,
        month = may,
       volume = {512},
       number = {3},
        pages = {3703-3716},
          doi = {10.1093/mnras/stac648},
archivePrefix = {arXiv},
       eprint = {2111.01160},
 primaryClass = {astro-ph.GA},
       adsurl = {https://ui.adsabs.harvard.edu/abs/2022MNRAS.512.3703B}
}

@ARTICLE{Ni+2022,
       author = {{Ni}, Yueying and {Di Matteo}, Tiziana and {Bird}, Simeon and {Croft}, Rupert and {Feng}, Yu and {Chen}, Nianyi and {Tremmel}, Michael and {DeGraf}, Colin and {Li}, Yin},
        title = "{The ASTRID simulation: the evolution of supermassive black holes}",
      journal = {\mnras},
         year = 2022,
        month = jun,
       volume = {513},
       number = {1},
        pages = {670-692},
          doi = {10.1093/mnras/stac351},
archivePrefix = {arXiv},
       eprint = {2110.14154},
 primaryClass = {astro-ph.GA},
       adsurl = {https://ui.adsabs.harvard.edu/abs/2022MNRAS.513..670N}
}

@ARTICLE{Weller2023,
       author = {{Weller}, Emma Jane and {Pacucci}, Fabio and {Ni}, Yueying and {Chen}, Nianyi and {Di Matteo}, Tiziana and {Siwek}, Magdalena and {Hernquist}, Lars},
        title = "{Orbital and radiative properties of wandering intermediate-mass black holes in the ASTRID simulation}",
      journal = {\mnras},
         year = 2023,
        month = apr,
       volume = {520},
       number = {3},
        pages = {3955-3963},
          doi = {10.1093/mnras/stad347},
archivePrefix = {arXiv},
       eprint = {2210.16319},
 primaryClass = {astro-ph.GA},
       adsurl = {https://ui.adsabs.harvard.edu/abs/2023MNRAS.520.3955W}
}

@ARTICLE{Chen+2022,
       author = {{Chen}, Nianyi and {Ni}, Yueying and {Tremmel}, Michael and {Di Matteo}, Tiziana and {Bird}, Simeon and {DeGraf}, Colin and {Feng}, Yu},
        title = "{Dynamical friction modelling of massive black holes in cosmological simulations and effects on merger rate predictions}",
      journal = {\mnras},
         year = 2022,
        month = feb,
       volume = {510},
       number = {1},
        pages = {531-550},
          doi = {10.1093/mnras/stab3411},
archivePrefix = {arXiv},
       eprint = {2104.00021},
 primaryClass = {astro-ph.GA},
       adsurl = {https://ui.adsabs.harvard.edu/abs/2022MNRAS.510..531C}
}

@ARTICLE{Ni+2025,
       author = {{Ni}, Yueying and {Chen}, Nianyi and {Zhou}, Yihao and {Park}, Minjung and {Yang}, Yanhui and {Di Matteo}, Tiziana and {Bird}, Simeon and {Croft}, Rupert},
        title = "{The Astrid Simulation: Evolution of Black Holes and Galaxies to z = 0.5 and Different Evolution Pathways for Galaxy Quenching}",
      journal = {\apj},
         year = 2025,
        month = sep,
       volume = {990},
       number = {2},
          eid = {120},
        pages = {120},
          doi = {10.3847/1538-4357/adf3a7},
archivePrefix = {arXiv},
       eprint = {2409.10666},
 primaryClass = {astro-ph.GA},
       adsurl = {https://ui.adsabs.harvard.edu/abs/2025ApJ...990..120N}
}

@software{Feng+2018,
       author = {{Feng}, Yu and {Bird}, Simeon and {Anderson}, Lauren and {Font-Ribera}, Andreu and {Pedersen}, Chris},
        title = "{MP-Gadget/MP-Gadget: A tag for getting a DOI}",
         year = 2018,
        month = oct,
          eid = {10.5281/zenodo.1451799},
          doi = {10.5281/zenodo.1451799},
      version = {FirstDOI},
    publisher = {Zenodo},
       adsurl = {https://ui.adsabs.harvard.edu/abs/2018zndo...1451799F}
}

@ARTICLE{Planck2020,
       author = {{Planck Collaboration} and {Aghanim}, N. and {Akrami}, Y. and {Ashdown}, M. and {Aumont}, J. and {Baccigalupi}, C. and {Ballardini}, M. and {Banday}, A.~J. and {Barreiro}, R.~B. and {Bartolo}, N. and {Basak}, S. and {Battye}, R. and {Benabed}, K. and {Bernard}, J.-P. and {Bersanelli}, M. and {Bielewicz}, P. and {Bock}, J.~J. and {Bond}, J.~R. and {Borrill}, J. and {Bouchet}, F.~R. and {Boulanger}, F. and {Bucher}, M. and {Burigana}, C. and {Butler}, R.~C. and {Calabrese}, E. and {Cardoso}, J.-F. and {Carron}, J. and {Challinor}, A. and {Chiang}, H.~C. and {Chluba}, J. and {Colombo}, L.~P.~L. and {Combet}, C. and {Contreras}, D. and {Crill}, B.~P. and {Cuttaia}, F. and {de Bernardis}, P. and {de Zotti}, G. and {Delabrouille}, J. and {Delouis}, J.-M. and {Di Valentino}, E. and {Diego}, J.~M. and {Dor{\'e}}, O. and {Douspis}, M. and {Ducout}, A. and {Dupac}, X. and {Dusini}, S. and {Efstathiou}, G. and {Elsner}, F. and {En{\ss}lin}, T.~A. and {Eriksen}, H.~K. and {Fantaye}, Y. and {Farhang}, M. and {Fergusson}, J. and {Fernandez-Cobos}, R. and {Finelli}, F. and {Forastieri}, F. and {Frailis}, M. and {Fraisse}, A.~A. and {Franceschi}, E. and {Frolov}, A. and {Galeotta}, S. and {Galli}, S. and {Ganga}, K. and {G{\'e}nova-Santos}, R.~T. and {Gerbino}, M. and {Ghosh}, T. and {Gonz{\'a}lez-Nuevo}, J. and {G{\'o}rski}, K.~M. and {Gratton}, S. and {Gruppuso}, A. and {Gudmundsson}, J.~E. and {Hamann}, J. and {Handley}, W. and {Hansen}, F.~K. and {Herranz}, D. and {Hildebrandt}, S.~R. and {Hivon}, E. and {Huang}, Z. and {Jaffe}, A.~H. and {Jones}, W.~C. and {Karakci}, A. and {Keih{\"a}nen}, E. and {Keskitalo}, R. and {Kiiveri}, K. and {Kim}, J. and {Kisner}, T.~S. and {Knox}, L. and {Krachmalnicoff}, N. and {Kunz}, M. and {Kurki-Suonio}, H. and {Lagache}, G. and {Lamarre}, J.-M. and {Lasenby}, A. and {Lattanzi}, M. and {Lawrence}, C.~R. and {Le Jeune}, M. and {Lemos}, P. and {Lesgourgues}, J. and {Levrier}, F. and {Lewis}, A. and {Liguori}, M. and {Lilje}, P.~B. and {Lilley}, M. and {Lindholm}, V. and {L{\'o}pez-Caniego}, M. and {Lubin}, P.~M. and {Ma}, Y.-Z. and {Mac{\'\i}as-P{\'e}rez}, J.~F. and {Maggio}, G. and {Maino}, D. and {Mandolesi}, N. and {Mangilli}, A. and {Marcos-Caballero}, A. and {Maris}, M. and {Martin}, P.~G. and {Martinelli}, M. and {Mart{\'\i}nez-Gonz{\'a}lez}, E. and {Matarrese}, S. and {Mauri}, N. and {McEwen}, J.~D. and {Meinhold}, P.~R. and {Melchiorri}, A. and {Mennella}, A. and {Migliaccio}, M. and {Millea}, M. and {Mitra}, S. and {Miville-Desch{\^e}nes}, M.-A. and {Molinari}, D. and {Montier}, L. and {Morgante}, G. and {Moss}, A. and {Natoli}, P. and {N{\o}rgaard-Nielsen}, H.~U. and {Pagano}, L. and {Paoletti}, D. and {Partridge}, B. and {Patanchon}, G. and {Peiris}, H.~V. and {Perrotta}, F. and {Pettorino}, V. and {Piacentini}, F. and {Polastri}, L. and {Polenta}, G. and {Puget}, J.-L. and {Rachen}, J.~P. and {Reinecke}, M. and {Remazeilles}, M. and {Renzi}, A. and {Rocha}, G. and {Rosset}, C. and {Roudier}, G. and {Rubi{\~n}o-Mart{\'\i}n}, J.~A. and {Ruiz-Granados}, B. and {Salvati}, L. and {Sandri}, M. and {Savelainen}, M. and {Scott}, D. and {Shellard}, E.~P.~S. and {Sirignano}, C. and {Sirri}, G. and {Spencer}, L.~D. and {Sunyaev}, R. and {Suur-Uski}, A.-S. and {Tauber}, J.~A. and {Tavagnacco}, D. and {Tenti}, M. and {Toffolatti}, L. and {Tomasi}, M. and {Trombetti}, T. and {Valenziano}, L. and {Valiviita}, J. and {Van Tent}, B. and {Vibert}, L. and {Vielva}, P. and {Villa}, F. and {Vittorio}, N. and {Wandelt}, B.~D. and {Wehus}, I.~K. and {White}, M. and {White}, S.~D.~M. and {Zacchei}, A. and {Zonca}, A.},
        title = "{Planck 2018 results. VI. Cosmological parameters}",
      journal = {\aap},
         year = 2020,
        month = sep,
       volume = {641},
          eid = {A6},
        pages = {A6},
          doi = {10.1051/0004-6361/201833910},
archivePrefix = {arXiv},
       eprint = {1807.06209},
 primaryClass = {astro-ph.CO},
       adsurl = {https://ui.adsabs.harvard.edu/abs/2020A&A...641A...6P}
}

@ARTICLE{Davis+1985,
       author = {{Davis}, M. and {Efstathiou}, G. and {Frenk}, C.~S. and {White}, S.~D.~M.},
        title = "{The evolution of large-scale structure in a universe dominated by cold dark matter}",
      journal = {\apj},
         year = 1985,
        month = may,
       volume = {292},
        pages = {371-394},
          doi = {10.1086/163168},
       adsurl = {https://ui.adsabs.harvard.edu/abs/1985ApJ...292..371D}
}

@ARTICLE{Springel+2001,
       author = {{Springel}, Volker and {White}, Simon D.~M. and {Tormen}, Giuseppe and {Kauffmann}, Guinevere},
        title = "{Populating a cluster of galaxies - I. Results at z=0}",
      journal = {\mnras},
         year = 2001,
        month = dec,
       volume = {328},
       number = {3},
        pages = {726-750},
          doi = {10.1046/j.1365-8711.2001.04912.x},
archivePrefix = {arXiv},
       eprint = {astro-ph/0012055},
 primaryClass = {astro-ph},
       adsurl = {https://ui.adsabs.harvard.edu/abs/2001MNRAS.328..726S}
}

@ARTICLE{DiMatteo+2005,
       author = {{Di Matteo}, Tiziana and {Springel}, Volker and {Hernquist}, Lars},
        title = "{Energy input from quasars regulates the growth and activity of black holes and their host galaxies}",
      journal = {\nat},
         year = 2005,
        month = feb,
       volume = {433},
       number = {7026},
        pages = {604-607},
          doi = {10.1038/nature03335},
archivePrefix = {arXiv},
       eprint = {astro-ph/0502199},
 primaryClass = {astro-ph},
       adsurl = {https://ui.adsabs.harvard.edu/abs/2005Natur.433..604D}
}

@ARTICLE{Tremmel2015,
       author = {{Tremmel}, M. and {Governato}, F. and {Volonteri}, M. and {Quinn}, T.~R.},
        title = "{Off the beaten path: a new approach to realistically model the orbital decay of supermassive black holes in galaxy formation simulations}",
      journal = {\mnras},
         year = 2015,
        month = aug,
       volume = {451},
       number = {2},
        pages = {1868-1874},
          doi = {10.1093/mnras/stv1060},
archivePrefix = {arXiv},
       eprint = {1501.07609},
 primaryClass = {astro-ph.GA},
       adsurl = {https://ui.adsabs.harvard.edu/abs/2015MNRAS.451.1868T}
}

@ARTICLE{Haidar+2022,
       author = {{Haidar}, Houda and {Habouzit}, M{\'e}lanie and {Volonteri}, Marta and {Mezcua}, Mar and {Greene}, Jenny and {Neumayer}, Nadine and {Angl{\'e}s-Alc{\'a}zar}, Daniel and {Martin-Navarro}, Ignacio and {Hoyer}, Nils and {Dubois}, Yohan and {Dav{\'e}}, Romeel},
        title = "{The black hole population in low-mass galaxies in large-scale cosmological simulations}",
      journal = {\mnras},
         year = 2022,
        month = aug,
       volume = {514},
       number = {4},
        pages = {4912-4931},
          doi = {10.1093/mnras/stac1659},
archivePrefix = {arXiv},
       eprint = {2201.09888},
 primaryClass = {astro-ph.GA},
       adsurl = {https://ui.adsabs.harvard.edu/abs/2022MNRAS.514.4912H}
}

@ARTICLE{Ricarte+2021_origins,
       author = {{Ricarte}, Angelo and {Tremmel}, Michael and {Natarajan}, Priyamvada and {Zimmer}, Charlotte and {Quinn}, Thomas},
        title = "{Origins and demographics of wandering black holes}",
      journal = {\mnras},
         year = 2021,
        month = jun,
       volume = {503},
       number = {4},
        pages = {6098-6111},
          doi = {10.1093/mnras/stab866},
archivePrefix = {arXiv},
       eprint = {2103.12124},
 primaryClass = {astro-ph.GA},
       adsurl = {https://ui.adsabs.harvard.edu/abs/2021MNRAS.503.6098R}
}

@ARTICLE{DiMatteo2023,
       author = {{Di Matteo}, Tiziana and {Ni}, Yueying and {Chen}, Nianyi and {Croft}, Rupert and {Bird}, Simeon and {Pacucci}, Fabio and {Ricarte}, Angelo and {Tremmel}, Michael},
        title = "{A vast population of wandering and merging IMBHs at cosmic noon}",
      journal = {\mnras},
         year = 2023,
        month = oct,
       volume = {525},
       number = {1},
        pages = {1479-1497},
          doi = {10.1093/mnras/stad2198},
archivePrefix = {arXiv},
       eprint = {2210.14960},
 primaryClass = {astro-ph.GA},
       adsurl = {https://ui.adsabs.harvard.edu/abs/2023MNRAS.525.1479D}
}

@ARTICLE{Tremmel+2018,
       author = {{Tremmel}, Michael and {Governato}, Fabio and {Volonteri}, Marta and {Pontzen}, Andrew and {Quinn}, Thomas R.},
        title = "{Wandering Supermassive Black Holes in Milky-Way-mass Halos}",
      journal = {\apjl},
         year = 2018,
        month = apr,
       volume = {857},
       number = {2},
          eid = {L22},
        pages = {L22},
          doi = {10.3847/2041-8213/aabc0a},
archivePrefix = {arXiv},
       eprint = {1802.06783},
 primaryClass = {astro-ph.GA},
       adsurl = {https://ui.adsabs.harvard.edu/abs/2018ApJ...857L..22T}
}

@ARTICLE{Zou+2025,
       author = {{Zou}, Fan and {Gallo}, Elena and {Seth}, Anil C. and {Hodges-Kluck}, Edmund and {Ohlson}, David and {Treu}, Tommaso and {Baldassare}, Vivienne F. and {Brandt}, W.~N. and {Greene}, Jenny E. and {Madau}, Piero and {Nguyen}, Dieu D. and {Plotkin}, Richard M. and {Reines}, Amy E. and {Sesana}, Alberto and {Woo}, Jong-Hak and {Wu}, Jianfeng},
        title = "{Central Massive Black Holes Are Not Ubiquitous in Local Low-mass Galaxies}",
      journal = {\apj},
         year = 2025,
        month = oct,
       volume = {992},
       number = {2},
          eid = {176},
        pages = {176},
          doi = {10.3847/1538-4357/ae06a1},
archivePrefix = {arXiv},
       eprint = {2510.05252},
 primaryClass = {astro-ph.GA},
       adsurl = {https://ui.adsabs.harvard.edu/abs/2025ApJ...992..176Z}
}

@software{pandas_development_team_2024,
  author       = {{The pandas development team}},
  title        = "{pandas-dev/pandas: Pandas}",
  month        = sep,
  year         = 2024,
  publisher    = {Zenodo},
  version      = {v2.2.3},
  doi          = {10.5281/zenodo.13819579},
  url          = {https://doi.org/10.5281/zenodo.13819579},
}

@InProceedings{McKinney2010,
  author    = {{McKinney}, Wes},
  title     = "{Data Structures for Statistical Computing in Python}",
  booktitle = {Proceedings of the 9th Python in Science Conference},
  pages     = {56 - 61},
  year      = {2010},
  editor    = {{van der Walt}, St\'efan and {Millman}, Jarrod},
  doi       = {10.25080/Majora-92bf1922-00a}
}

@ARTICLE{Hunter_2007,
       author = {{Hunter}, John D.},
        title = "{Matplotlib: A 2D Graphics Environment}",
      journal = {Computing in Science and Engineering},
         year = 2007,
        month = jan,
       volume = {9},
       number = {3},
        pages = {90-95},
          doi = {10.1109/MCSE.2007.55},
       adsurl = {https://ui.adsabs.harvard.edu/abs/2007CSE.....9...90H}
}

@ARTICLE{Virtanen+2020,
       author = {{Virtanen}, Pauli and {Gommers}, Ralf and {Oliphant}, Travis E. and {Haberland}, Matt and {Reddy}, Tyler and {Cournapeau}, David and {Burovski}, Evgeni and {Peterson}, Pearu and {Weckesser}, Warren and {Bright}, Jonathan and {van der Walt}, St{\'e}fan J. and {Brett}, Matthew and {Wilson}, Joshua and {Millman}, K. Jarrod and {Mayorov}, Nikolay and {Nelson}, Andrew R.~J. and {Jones}, Eric and {Kern}, Robert and {Larson}, Eric and {Carey}, C.~J. and {Polat}, {\.I}lhan and {Feng}, Yu and {Moore}, Eric W. and {VanderPlas}, Jake and {Laxalde}, Denis and {Perktold}, Josef and {Cimrman}, Robert and {Henriksen}, Ian and {Quintero}, E.~A. and {Harris}, Charles R. and {Archibald}, Anne M. and {Ribeiro}, Ant{\^o}nio H. and {Pedregosa}, Fabian and {van Mulbregt}, Paul and {SciPy 1. 0 Contributors}},
        title = "{SciPy 1.0: fundamental algorithms for scientific computing in Python}",
      journal = {Nature Methods},
         year = 2020,
        month = feb,
       volume = {17},
        pages = {261-272},
          doi = {10.1038/s41592-019-0686-2},
archivePrefix = {arXiv},
       eprint = {1907.10121},
 primaryClass = {cs.MS},
       adsurl = {https://ui.adsabs.harvard.edu/abs/2020NatMe..17..261V}
}

@Article{         Harris+2020,
 title         = {Array programming with {NumPy}},
 author        = {Charles R. Harris and K. Jarrod Millman and St{\'{e}}fan J.
                 van der Walt and Ralf Gommers and Pauli Virtanen and David
                 Cournapeau and Eric Wieser and Julian Taylor and Sebastian
                 Berg and Nathaniel J. Smith and Robert Kern and Matti Picus
                 and Stephan Hoyer and Marten H. van Kerkwijk and Matthew
                 Brett and Allan Haldane and Jaime Fern{\'{a}}ndez del
                 R{\'{i}}o and Mark Wiebe and Pearu Peterson and Pierre
                 G{\'{e}}rard-Marchant and Kevin Sheppard and Tyler Reddy and
                 Warren Weckesser and Hameer Abbasi and Christoph Gohlke and
                 Travis E. Oliphant},
 year          = {2020},
 month         = sep,
 journal       = {Nature},
 volume        = {585},
 number        = {7825},
 pages         = {357-362},
 doi           = {10.1038/s41586-020-2649-2},
 publisher     = {Springer Science and Business Media {LLC}},
 url           = {https://doi.org/10.1038/s41586-020-2649-2}
}

@article{astropy:2013,
Adsurl = {http://adsabs.harvard.edu/abs/2013A%26A...558A..33A},
Archiveprefix = {arXiv},
Author = {{Astropy Collaboration} and {Robitaille}, T.~P. and {Tollerud}, E.~J. and {Greenfield}, P. and {Droettboom}, M. and {Bray}, E. and {Aldcroft}, T. and {Davis}, M. and {Ginsburg}, A. and {Price-Whelan}, A.~M. and {Kerzendorf}, W.~E. and {Conley}, A. and {Crighton}, N. and {Barbary}, K. and {Muna}, D. and {Ferguson}, H. and {Grollier}, F. and {Parikh}, M.~M. and {Nair}, P.~H. and {Unther}, H.~M. and {Deil}, C. and {Woillez}, J. and {Conseil}, S. and {Kramer}, R. and {Turner}, J.~E.~H. and {Singer}, L. and {Fox}, R. and {Weaver}, B.~A. and {Zabalza}, V. and {Edwards}, Z.~I. and {Azalee Bostroem}, K. and {Burke}, D.~J. and {Casey}, A.~R. and {Crawford}, S.~M. and {Dencheva}, N. and {Ely}, J. and {Jenness}, T. and {Labrie}, K. and {Lim}, P.~L. and {Pierfederici}, F. and {Pontzen}, A. and {Ptak}, A. and {Refsdal}, B. and {Servillat}, M. and {Streicher}, O.},
Doi = {10.1051/0004-6361/201322068},
Eid = {A33},
Eprint = {1307.6212},
Journal = {\aap},
Month = oct,
Pages = {A33},
Primaryclass = {astro-ph.IM},
Title = {{Astropy: A community Python package for astronomy}},
Volume = 558,
Year = 2013}

@ARTICLE{astropy:2018,
       author = {{Astropy Collaboration} and {Price-Whelan}, A.~M. and
         {Sip{\H{o}}cz}, B.~M. and {G{\"u}nther}, H.~M. and {Lim}, P.~L. and
         {Crawford}, S.~M. and {Conseil}, S. and {Shupe}, D.~L. and
         {Craig}, M.~W. and {Dencheva}, N. and {Ginsburg}, A. and {Vand
        erPlas}, J.~T. and {Bradley}, L.~D. and {P{\'e}rez-Su{\'a}rez}, D. and
         {de Val-Borro}, M. and {Aldcroft}, T.~L. and {Cruz}, K.~L. and
         {Robitaille}, T.~P. and {Tollerud}, E.~J. and {Ardelean}, C. and
         {Babej}, T. and {Bach}, Y.~P. and {Bachetti}, M. and {Bakanov}, A.~V. and
         {Bamford}, S.~P. and {Barentsen}, G. and {Barmby}, P. and
         {Baumbach}, A. and {Berry}, K.~L. and {Biscani}, F. and {Boquien}, M. and
         {Bostroem}, K.~A. and {Bouma}, L.~G. and {Brammer}, G.~B. and
         {Bray}, E.~M. and {Breytenbach}, H. and {Buddelmeijer}, H. and
         {Burke}, D.~J. and {Calderone}, G. and {Cano Rodr{\'\i}guez}, J.~L. and
         {Cara}, M. and {Cardoso}, J.~V.~M. and {Cheedella}, S. and {Copin}, Y. and
         {Corrales}, L. and {Crichton}, D. and {D'Avella}, D. and {Deil}, C. and
         {Depagne}, {\'E}. and {Dietrich}, J.~P. and {Donath}, A. and
         {Droettboom}, M. and {Earl}, N. and {Erben}, T. and {Fabbro}, S. and
         {Ferreira}, L.~A. and {Finethy}, T. and {Fox}, R.~T. and
         {Garrison}, L.~H. and {Gibbons}, S.~L.~J. and {Goldstein}, D.~A. and
         {Gommers}, R. and {Greco}, J.~P. and {Greenfield}, P. and
         {Groener}, A.~M. and {Grollier}, F. and {Hagen}, A. and {Hirst}, P. and
         {Homeier}, D. and {Horton}, A.~J. and {Hosseinzadeh}, G. and {Hu}, L. and
         {Hunkeler}, J.~S. and {Ivezi{\'c}}, {\v{Z}}. and {Jain}, A. and
         {Jenness}, T. and {Kanarek}, G. and {Kendrew}, S. and {Kern}, N.~S. and
         {Kerzendorf}, W.~E. and {Khvalko}, A. and {King}, J. and {Kirkby}, D. and
         {Kulkarni}, A.~M. and {Kumar}, A. and {Lee}, A. and {Lenz}, D. and
         {Littlefair}, S.~P. and {Ma}, Z. and {Macleod}, D.~M. and
         {Mastropietro}, M. and {McCully}, C. and {Montagnac}, S. and
         {Morris}, B.~M. and {Mueller}, M. and {Mumford}, S.~J. and {Muna}, D. and
         {Murphy}, N.~A. and {Nelson}, S. and {Nguyen}, G.~H. and
         {Ninan}, J.~P. and {N{\"o}the}, M. and {Ogaz}, S. and {Oh}, S. and
         {Parejko}, J.~K. and {Parley}, N. and {Pascual}, S. and {Patil}, R. and
         {Patil}, A.~A. and {Plunkett}, A.~L. and {Prochaska}, J.~X. and
         {Rastogi}, T. and {Reddy Janga}, V. and {Sabater}, J. and
         {Sakurikar}, P. and {Seifert}, M. and {Sherbert}, L.~E. and
         {Sherwood-Taylor}, H. and {Shih}, A.~Y. and {Sick}, J. and
         {Silbiger}, M.~T. and {Singanamalla}, S. and {Singer}, L.~P. and
         {Sladen}, P.~H. and {Sooley}, K.~A. and {Sornarajah}, S. and
         {Streicher}, O. and {Teuben}, P. and {Thomas}, S.~W. and
         {Tremblay}, G.~R. and {Turner}, J.~E.~H. and {Terr{\'o}n}, V. and
         {van Kerkwijk}, M.~H. and {de la Vega}, A. and {Watkins}, L.~L. and
         {Weaver}, B.~A. and {Whitmore}, J.~B. and {Woillez}, J. and
         {Zabalza}, V. and {Astropy Contributors}},
        title = "{The Astropy Project: Building an Open-science Project and Status of the v2.0 Core Package}",
      journal = {\aj},
         year = 2018,
        month = sep,
       volume = {156},
       number = {3},
          eid = {123},
        pages = {123},
          doi = {10.3847/1538-3881/aabc4f},
archivePrefix = {arXiv},
       eprint = {1801.02634},
 primaryClass = {astro-ph.IM},
       adsurl = {https://ui.adsabs.harvard.edu/abs/2018AJ....156..123A}
}

@ARTICLE{Sharma+2022,
       author = {{Sharma}, Ray S. and {Brooks}, Alyson M. and {Tremmel}, Michael and {Bellovary}, Jillian and {Ricarte}, Angelo and {Quinn}, Thomas R.},
        title = "{A Hidden Population of Massive Black Holes in Simulated Dwarf Galaxies}",
      journal = {\apj},
         year = 2022,
        month = sep,
       volume = {936},
       number = {1},
          eid = {82},
        pages = {82},
          doi = {10.3847/1538-4357/ac8664},
archivePrefix = {arXiv},
       eprint = {2203.05580},
 primaryClass = {astro-ph.GA},
       adsurl = {https://ui.adsabs.harvard.edu/abs/2022ApJ...936...82S}
}

@ARTICLE{Bhowmick2025,
       author = {{Bhowmick}, Aklant K. and {Blecha}, Laura and {Torrey}, Paul and {Somerville}, Rachel S. and {Kelley}, Luke Zoltan and {Weinberger}, Rainer and {Vogelsberger}, Mark and {Hernquist}, Lars and {Natarajan}, Priyamvada and {Kho}, Jonathan and {Di Matteo}, Tiziana},
        title = "{Signatures of black hole seeding in the local Universe: predictions from the BRAHMA cosmological simulations}",
      journal = {\mnras},
         year = 2025,
        month = mar,
       volume = {538},
       number = {1},
        pages = {518-536},
          doi = {10.1093/mnras/staf269},
archivePrefix = {arXiv},
       eprint = {2411.19332},
 primaryClass = {astro-ph.GA},
       adsurl = {https://ui.adsabs.harvard.edu/abs/2025MNRAS.538..518B}
}

@ARTICLE{Miller2015,
       author = {{Miller}, Brendan P. and {Gallo}, Elena and {Greene}, Jenny E. and {Kelly}, Brandon C. and {Treu}, Tommaso and {Woo}, Jong-Hak and {Baldassare}, Vivienne},
        title = "{X-Ray Constraints on the Local Supermassive Black Hole Occupation Fraction}",
      journal = {\apj},
         year = 2015,
        month = jan,
       volume = {799},
       number = {1},
          eid = {98},
        pages = {98},
          doi = {10.1088/0004-637X/799/1/98},
archivePrefix = {arXiv},
       eprint = {1403.4246},
 primaryClass = {astro-ph.GA},
       adsurl = {https://ui.adsabs.harvard.edu/abs/2015ApJ...799...98M}
}

@ARTICLE{Tremmel2024,
       author = {{Tremmel}, Michael and {Ricarte}, Angelo and {Natarajan}, Priyamvada and {Bellovary}, Jillian and {Sharma}, Ray and {Quinn}, Thomas R.},
        title = "{An Enhanced Massive Black Hole Occupation Fraction Predicted in Cluster Dwarf Galaxies}",
      journal = {The Open Journal of Astrophysics},
         year = 2024,
        month = apr,
       volume = {7},
          eid = {26},
        pages = {26},
          doi = {10.33232/001c.116617},
archivePrefix = {arXiv},
       eprint = {2306.12813},
 primaryClass = {astro-ph.GA},
       adsurl = {https://ui.adsabs.harvard.edu/abs/2024OJAp....7E..26T}
}

@ARTICLE{Contini2026,
       author = {{Contini}, Emanuele and {Jang}, J.~K. and {Rhee}, Jinsu and {Seo}, Changjo and {Yi}, Sukyoung K.},
        title = "{Black Hole Occupation Fraction: Dependence on Black Hole Mass Threshold, Environment, Resolution and Redshift}",
      journal = {arXiv e-prints},
         year = 2026,
        month = jun,
          eid = {arXiv:2606.22758},
        pages = {arXiv:2606.22758},
          doi = {10.48550/arXiv.2606.22758},
archivePrefix = {arXiv},
       eprint = {2606.22758},
 primaryClass = {astro-ph.GA},
       adsurl = {https://ui.adsabs.harvard.edu/abs/2026arXiv260622758C}
}

@software{WebPlotDigitizer,
    author = {Ankit Rohatgi},
    title = {WebPlotDigitizer},
    url = {https://automeris.io},
    version = {5.2},
    year = 2024
}
\bibliographystyle{aasjournalv7}

\end{document}